\documentclass[twocolumn]{jpsj3}
\usepackage{txfonts}
\usepackage{color}

\title{Magnetic Impurity Effects on Ferromagnetic Fluctuations \\  in Heavily Overdoped (Bi,Pb)$_{2}$Sr$_{2}$Cu$_{1-y}$Fe$_{y}$O$_{6+\delta}$ Cuprates}

\author{Yota Komiyama$^1$, Shusei Onishi$^1$, Mayuko Harada$^1$, Hideki Kuwahara$^1$, Haruhiko Kuroe$^1$, Koshi Kurashima$^2$,Takayuki Kawamata$^2$, Yoji Koike$^2$, Isao Watanabe$^3$, and Tadashi Adachi$^1$\thanks{E-mail: t-adachi@sophia.ac.jp}}
\inst{$^1$Department of Engineering and Applied Sciences, Sophia University, 7-1 Kioi-cho, Chiyoda-ku, Tokyo 102-8554, Japan \\
$^2$Department of Applied Physics, Tohoku University, 6-6-05 Aoba, Aramaki, Sendai 980-8579, Japan \\
$^3$Meson Science Laboratory, Nishina Center for Accelerator-Based Science, RIKEN, 2-1 Hirosawa, Wako 351-0198, Japan}

\abst{We investigated the effects of Fe substitution on ferromagnetic fluctuations in the superconducting overdoped and non-superconducting heavily overdoped regimes of Bi-2201 cuprates by magnetization and electrical resistivity measurements.
It was found that the spin-glass state was induced at low temperatures by the Fe substitution.
The Curie constant and the effective Bohr magneton, estimated from magnetic susceptibility, as well as the dimensionality of the ferromagnetic fluctuations from the resistivity, suggest the enhancement of the ferromagnetic fluctuations caused by the Fe substitution.
A ferromagnetic spin-cluster model is proposed for the heavily overdoped regime of Bi-2201, whereas an antiferromagnetic spin-cluster model has been proposed for the overdoped regime of Bi-2201 [Hiraka {\it et al}., Phys. Rev. {\bf B} 81, 144501 (2010)].}

\kword{Bi-2201 cuprate, ferromagnetic fluctuations, heavily overdoped regime, magnetic impurity}

\begin{document}
\maketitle

\section{Introduction}

In the high-$T_{\rm c}$ cuprate superconductivity, the relationship between antiferromagnetic (AF) spin fluctuations and superconductivity has been extensively studied.
In the overdoped regime of the hole-doped cuprates of La$_{2-x}$Sr$_{x}$CuO$_{4}$, in which the superconducting (SC) transition temperature $T_{\rm c}$ is depressed by hole doping, inelastic neutron scattering experiments  revealed that low-energy incommensurate AF fluctuations vanish concomitant with the suppression of superconductivity.~\cite{Swakimoto}
The energy at which the dynamic spin susceptibility $\chi^{\prime \prime}$($\omega$) is maximum is unchanged with hole doping, suggesting that a phase separation into SC and non-SC regions takes place.~\cite{YJuemura,Ytanabe}
The suppression of AF fluctuations by overdoping was also observed by muon-spin-relaxation ($\mu$SR) measurements in Zn-substituted La$_{2-x}$Sr$_{x}$Cu$_{1-y}$Zn$_{y}$O$_{4}$ \cite{Resdiana} and Bi-2201 of Bi$_{1.74}$Pb$_{0.38}$Sr$_{1.88}$Cu$_{1-y}$Zn$_{y}$O$_{6+\delta}$.~\cite{TadachiPRB}
On the other hand, resonant inelastic X-ray scattering (RIXS) experiments revealed that high-energy AF fluctuations are robust in the non-SC heavily overdoped (HOD) regime of La$_{2-x}$Sr$_{x}$CuO$_{4}$.~\cite{MPMdean} 
Recently, a charge order, generally competing with the superconductivity, has been observed in non-SC HOD Bi-2201, \cite{Xli,YYpengCONat} as well as at the hole concentration of $\sim$1/8 per Cu. \cite{YYpengCO,Skawasaki}
These suggest that the weakening of AF fluctuations in the overdoped regime is not only the cause of the suppression of superconductivity.

Several experimental and theoretical works suggest that the suppression of superconductivity in the HOD regime is related to the presence of ferromagnetic order/fluctuations.
The quantum critical scaling theory for HOD Tl$_{2}$Ba$_{2}$CuO$_{6+\delta}$ suggests that the enhancement of the magnetic susceptibility $\chi$ at low temperatures is caused by ferromagnetic fluctuations due to itinerant electrons.~\cite{Akopp}
The development of ferromagnetic fluctuations in HOD cuprates was also suggested from the theoretical calculations in which the spectral weight of the spin dynamical structure factor \cite{CJjia} and the spin susceptibility \cite{Steranishi,TAmaier} increased at around $Q = (0,0)$.
Experimentally, a three-dimensional (3D) ferromagnetic order due to itinerant electrons was suggested in HOD La$_{2-x}$Sr$_{x}$CuO$_{4}$ from $\mu$SR and transport measurements.~\cite{JEsonier}
From our previous study on HOD Bi-2201 \cite{Kkurashima}, $\mu$SR measurements revealed the development of spin fluctuations at low temperatures.
The resistivity, $\chi$ and specific heat exhibited behaviors characteristic of two-dimensional (2D) ferromagnetic fluctuations according to the self-consistent renormalization (SCR) theory~\cite{Yhatatani}.
In Bi-2201, the RIXS results revealed that the spectral weight was redistributed from AF $Q = (0.5,0.5)$ in the overdoped regime to ferromagnetic $Q = (0,0)$ in the HOD regime.~\cite{YYpengAF}
Quite recently, a ferromagnetic order has been proposed for the HOD regime of the electron-doped cuprates La$_{2-x}$Ce$_{x}$CuO$_{4}$.~\cite{Tsakar}

Studies of impurity effects are useful for clarifying the nature of spin fluctuations.
In the overdoped regime of La$_{2-x}$Sr$_{x}$Cu$_{1-y}$Fe$_{y}$O$_{4}$ in which a magnetic impurity Fe is substituted for Cu, neutron scattering and angle-resolved photoemission spectroscopy measurements suggest the formation of spin density waves due to the Fermi-surface nesting.~\cite{RHhe}
The $\mu$SR and $\chi$ measurements of overdoped La$_{2-x}$Sr$_{x}$Cu$_{1-y}$Fe$_{y}$O$_{4}$ revealed the formation of a spin-glass state due to the Ruderman-Kittel-Kasuya-Yosida (RKKY) interaction between Fe$^{3+}$ spins.~\cite{KMsuzuki}
Hiraka ${\it et \ al}$. suggested on the basis of experimental results on elastic neutron scattering in Fe-substituted Bi$_{1.75}$Pb$_{0.35}$Sr$_{1.90}$Cu$_{1-y}$Fe$_{y}$O$_{6+\delta}$ that AF spin clusters are formed around Fe.~\cite{Hhiraka}
In non-SC HOD Bi-2201, our previous $\mu$SR measurements revealed that the 5$\%$ Fe substitution for Cu enhanced the spin correlation and induced some forms of spin ordering at low temperatures, suggesting the development of ferromagnetic fluctuations caused by the Fe substitution.~\cite{Tadachi}

In this paper, we report the effects of Fe substitution on ferromagnetic fluctuations in SC overdoped and non-SC HOD (Bi,Pb)$_{2}$Sr$_{2}$Cu$_{1-y}$Fe$_{y}$O$_{6+\delta}$ at various Fe contents.
We found that $\chi$ exhibited the hysteresis between field cooling (FC) and zero-field cooling (ZFC) at low temperatures owing to the formation of a spin-glass state.
The onset temperature of the hysteresis increases with increasing Fe content, suggesting the Fe-induced spin glass formation due to the RKKY interaction between Fe$^{3+}$ spins.
The Curie constant obtained from $\chi$ increases in proportion to the Fe content, and the estimated effective Bohr magneton exceeded that of a bare Fe$^{3+}$ spin. 
These results suggest that the ferromagnetic fluctuations develop owing to the Fe substitution.
We propose the formation of ferromagnetic spin clusters around Fe in HOD Bi-2201, as opposed to that of AF spin clusters around Fe in the overdoped regime.~\cite{Hhiraka}

\section{Experimental Procedure}

Single crystals of Bi$_{1.74}$Pb$_{0.38}$Sr$_{1.88}$Cu$_{1-y}$Fe$_{y}$O$_{6+\delta}$ ($y$ = 0, 0.03, 0.05, 0.09) were grown by the floating-zone method.~\cite{Kkurashima}
The obtained crystals were confirmed to be of the single phase by X-ray diffraction analysis. 
The Fe content $y$ was determined by inductively coupled plasma (ICP) analysis.
The distribution of the Fe content in a sample was investigated by electron probe microanalysis (EPMA).
From the ICP analysis and EPMA results, the Fe contents were determined to be $y$ = 0, 0.03(0), 0.05(0), and 0.096(10).
As-grown samples of $y$ = 0 exhibit superconductivity at $T_{\rm c} = 5 \ {\rm K}$ and therefore reside in the overdoped regime \cite{Kkurashima}, whereas samples in the HOD regime, where the superconductivity disappears at $y$ = 0, were obtained by annealing in flowing oxygen at 500 $^\circ$C for 24 h.
Magnetization was measured using a SC quantum interference device magnetometer (Quantum Design, MPMS-XL5, -XL7). Electrical resistivity was measured by the standard dc four-probe method using a commercial apparatus (Quantum Design, PPMS).

\section{Results}

\begin{figure}[tbp]
\includegraphics[width=1.0\linewidth]{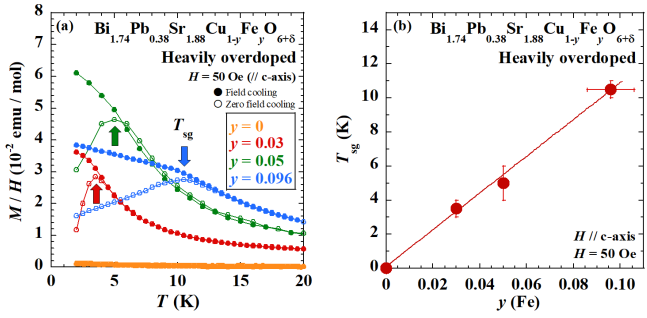}
\caption{(Color online) (a) Temperature dependence of the magnetic susceptibility in a magnetic field of 50 Oe parallel to the c-axis on zero-field cooling (ZFC) and field cooling (FC) for Bi$_{1.74}$Pb$_{0.38}$Sr$_{1.88}$Cu$_{1-y}$Fe$_{y}$O$_{6+\delta}$ with $y = 0${--}$0.096$. Arrows represent $T_{\rm sg}$ defined as the peak temperature of the magnetic susceptibility on ZFC. (b) Fe concentration dependence of $T_{\rm sg}$ for Bi$_{1.74}$Pb$_{0.38}$Sr$_{1.88}$Cu$_{1-y}$Fe$_{y}$O$_{6+\delta}$ with $y = 0${--}$0.096$. Solid lines are guides to the eye.}
\label{fig:f1}
\end{figure}

Figure 1(a) shows the temperature dependence of $\chi$ on ZFC and FC in HOD Bi$_{1.74}$Pb$_{0.38}$Sr$_{1.88}$Cu$_{1-y}$Fe$_{y}$O$_{6+\delta}$ at ${\it y} = 0${--}$0.096$.
A hysteresis is observed for Fe-substituted samples.
Moreover, we found that the peak temperature of $\chi$ in ZFC $T_{\rm sg}$, shown by arrows, increases with the Fe content.
The hysteresis behavior is similar to that observed in Fe-substituted Bi-2201 in the overdoped regime \cite{Hhiraka}, suggesting the formation of a spin-glass state.
The Fe concentration dependence of $T_{\rm sg}$ is shown in Fig. 1(b).
The absence of $T_{\rm sg}$ at ${\it y} = 0$ and a linear relationship between $T_{\rm sg}$ and the Fe content suggest that the formation of a spin-glass state is related to the substituted Fe.
\begin{figure*}[tbp]
\begin{center}
\includegraphics[width=0.67\linewidth]{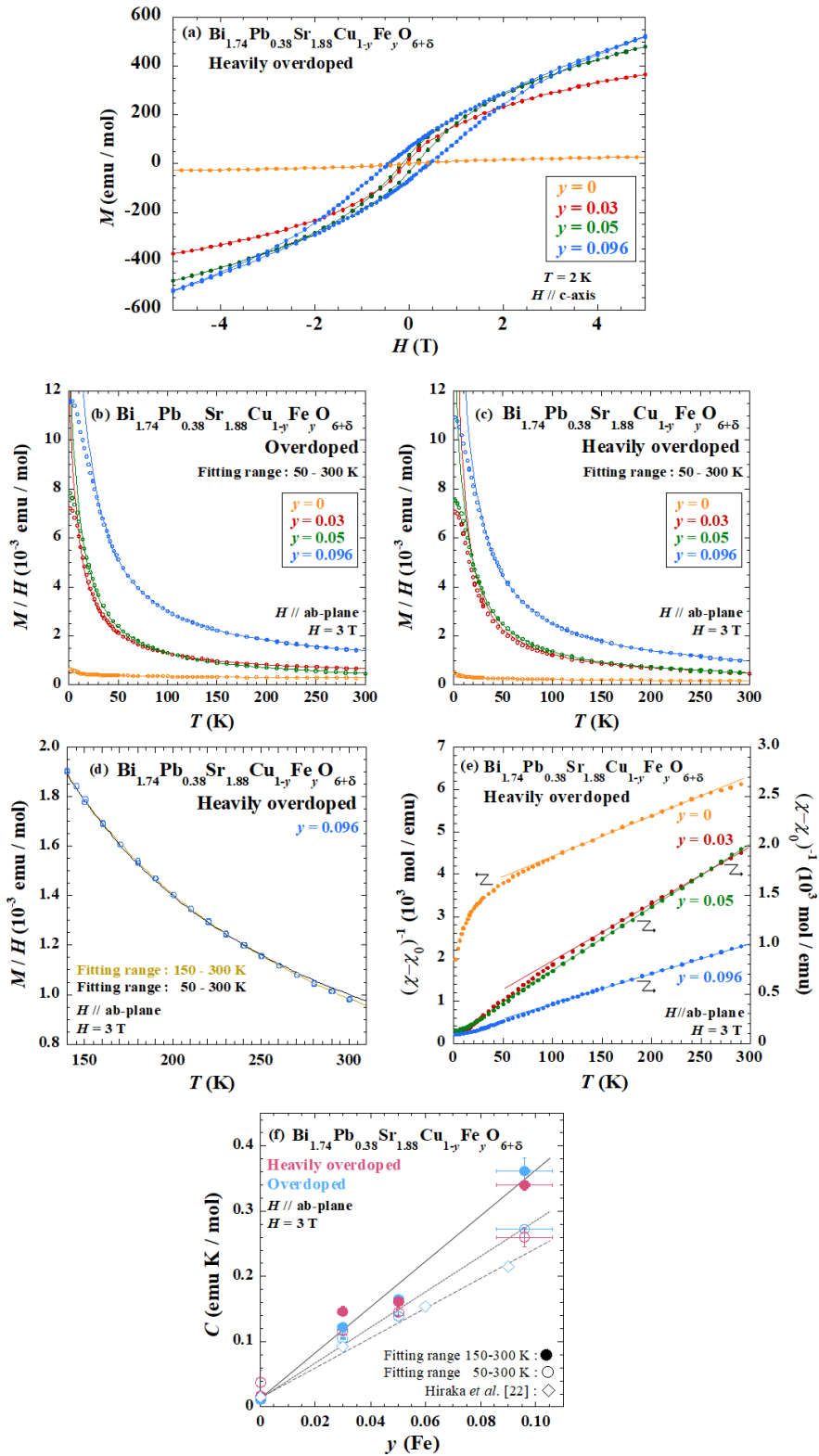}
\caption{(Color online) (a) Magnetization curves of HOD Bi$_{1.74}$Pb$_{0.38}$Sr$_{1.88}$Cu$_{1-y}$Fe$_{y}$O$_{6+\delta}$ with $y = 0${--}$0.096$ in a magnetic field along the c-axis. (b)(c) Temperature dependence of the magnetic susceptibility in a magnetic field at 3 T parallel to the ab-plane in overdoped and HOD Bi$_{1.74}$Pb$_{0.38}$Sr$_{1.88}$Cu$_{1-y}$Fe$_{y}$O$_{6+\delta}$ with $y = 0${--}$0.096$. Solid lines are the best-fit results in the fitting range of 50--300 K obtained by Eq. (1). (d) Temperature dependence of the magnetic susceptibility in HOD Bi$_{1.74}$Pb$_{0.38}$Sr$_{1.88}$Cu$_{1-y}$Fe$_{y}$O$_{6+\delta}$ with $y = 0.096$ with the fitting results in the ranges of 50--300 K and 150--300 K. (e) Temperature dependence of the inverse susceptibility ($\chi-\chi_0$)$^{-1}$ in HOD Bi$_{1.74}$Pb$_{0.38}$Sr$_{1.88}$Cu$_{1-y}$Fe$_{y}$O$_{6+\delta}$ with $y = 0${--}$0.096$. Solid lines are guides to the eye.  (f) Fe concentration dependence of the Curie constant obtained by fitting the magnetic susceptibility in (b) and (c) in the ranges of 50--300 K and 150--300 K. The overdoped data obtained by Hiraka $et \ al$. \cite{Hhiraka} are also plotted for comparison. Solid lines are guides to the eye.}
\label{fig:f2}
\end{center}
\end{figure*}

Figure 2(a) shows the magnetization curves at 2 K for HOD Bi$_{1.74}$Pb$_{0.38}$Sr$_{1.88}$Cu$_{1-y}$Fe$_{y}$O$_{6+\delta}$ with ${\it y} = 0${--}$0.096$.
Although the magnetization is almost linear in the magnetic field for ${\it y} = 0$, it tends to saturate in high magnetic fields at ${\it y} = 0.03${--}$0.096$.
Moreover, a hysteresis loop is clearly observed at $y = 0.05 \ \rm{and} \ 0.096$, corresponding to the hysteresis in the temperature dependence shown in Fig. 1(a).
Figures 2(b) and 2(c) show the temperature dependence of $\chi$ in a magnetic field at 3 T for overdoped and HOD Bi$_{1.74}$Pb$_{0.38}$Sr$_{1.88}$Cu$_{1-y}$Fe$_{y}$O$_{6+\delta}$ with ${\it y} = 0${--}$0.096$, respectively.
For impurity-free ${\it y} = 0$, $\chi$ is proportional to $T$ln$T$ at low temperatures, suggesting 2D ferromagnetic fluctuations.~\cite{Yhatatani}
For both overdoped and HOD Bi-2201, $\chi$ increases with increasing Fe content.
To investigate the effects of Fe spins in detail, the data were fitted on the basis of the Curie--Weiss law shown in Eq. (1).
\begin{equation}
\chi = \chi_0  + C/(T + \Theta)
\end{equation}
$\chi_0$ is the temperature-independent term, $C$ the Curie constant, and $\Theta$ the Weiss temperature. The fitting range was set to be 50--300 K, in which the magnetization is proportional to the magnetic field, to eliminate the effects of the hysteresis at low temperatures.
As shown in Figs. 2(b) and 2(c), the data fitted on the basis of the Curie--Weiss law are roughly reproduced.
Focusing on the high-temperature range in Fig. 2(d), however, we see that the fitting line deviates from the data.
The temperature dependence of the inverse susceptibility ($\chi-\chi_0$)$^{-1}$ of HOD Bi$_{1.74}$Pb$_{0.38}$Sr$_{1.88}$Cu$_{1-y}$Fe$_{y}$O$_{6+\delta}$ with ${\it y} = 0${--}$0.096$ is shown in Fig. 2(e).
We found that ($\chi-\chi_0$)$^{-1}$ below $\sim$ 150 K deviates from the linear relationship at high temperatures. These suggest that the magnetic interaction to form a spin cluster at low temperatures (described later) appears below $\sim$ 150 K.
Therefore, we also performed fitting in the high-temperature range of 150--300 K, so that the fitting clearly becomes better.
Figure 2(f) shows the Fe concentration dependence of the Curie constant, obtained in fitting ranges of 50--300 K and 150--300 K, for overdoped and HOD Bi$_{1.74}$Pb$_{0.38}$Sr$_{1.88}$Cu$_{1-y}$Fe$_{y}$O$_{6+\delta}$ with $y = 0${--}$0.096$, together with the Curie constant reported by Hiraka $et \ al$.~\cite{Hhiraka}
The Curie constant increases with increasing $y$ for both fitting ranges of 50--300 K and 150--300 K.
For the fitting range of 50--300 K, the Curie constant is almost the same between the overdoped and HOD regimes and close to that reported by Hiraka $et \ al$.~\cite{Hhiraka}
In the fitting range of 150--300 K, however, we found that the Curie constant values are larger than those obtained in the range of 50--300 K.
The effective Bohr magneton $p_{\rm eff}$ was calculated using $C$=($N$${\mu_B^2}$/3$k_B$)$p_{\rm eff}^2$ for HOD Bi$_{1.74}$Pb$_{0.38}$Sr$_{1.88}$Cu$_{1-y}$Fe$_{y}$O$_{6+\delta}$ with $y = 0.096$.
The $p_{\rm eff}$ in the range of 150--300 K for HOD $y = 0.096$ is 6.2, which is larger than 5.9 for a bare Fe$^{3+}$ spin (s = 5/2).
On the other hand, $p_{\rm eff}$ in the range of 50--300 K for HOD $y = 0.096$ is 5.4.

\begin{figure}[tbp]
\includegraphics[width=1.0\linewidth]{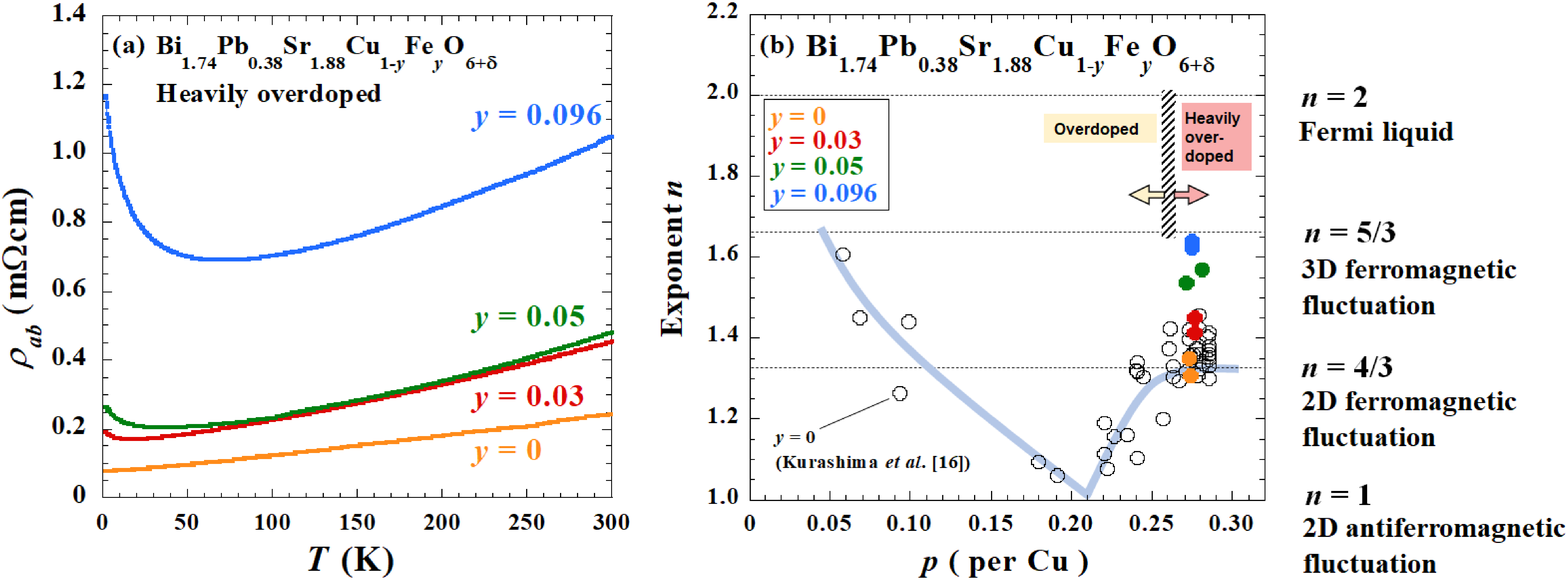}
\caption{(Color online) (a) Temperature dependence of the ab-plane electrical resistivity $\rho_{\rm ab}$ for HOD Bi$_{1.74}$Pb$_{0.38}$Sr$_{1.88}$Cu$_{1-y}$Fe$_{y}$O$_{6+\delta}$ with $y = 0${--}$0.096$. (b) Hole-concentration dependence of the exponent $n$ of $\rho_{\rm ab}$ = $\rho_{0}$ + A$T^{n}
$ for Bi$_{1.74}$Pb$_{0.38}$Sr$_{1.88}$Cu$_{1-y}$Fe$_{y}$O$_{6+\delta}$ with $y = 0${--}$0.096$. Open circles are the data of impurity-free $y = 0$.\cite{Kkurashima} }
\label{fig:f3}
\end{figure}

Figure 3(a) shows the temperature dependence of the ab-plane electrical resistivity $\rho_{\rm ab}$ for HOD Bi$_{1.74}$Pb$_{0.38}$Sr$_{1.88}$Cu$_{1-y}$Fe$_{y}$O$_{6+\delta}$ with $y = 0${--}$0.096$.
No SC transition is observed in all HOD samples.
An upturn of $\rho_{\rm ab}$ is enhanced by the Fe substitution, suggesting the localization of carriers by magnetic impurities.
To investigate the characteristic behavior of $\rho_{ab}$ caused by the ferromagnetic fluctuations, $\rho_{\rm ab}$ was fitted using the following power-law equation, where $n$ is the exponent, $\rho_{0}$ is the temperature-independent term, and $A$ is the temperature coefficient.
\begin{equation}
\rho_{ab} = \rho_0+AT^n
\end{equation}
In general, $n$ is estimated from $\rho_{\rm ab}$ at low temperatures according to the SCR theory.
As shown in Fig. 3(a), however, the Fe substitution induces the localization of carriers.
As $\rho_{\rm ab}$ of $y = 0$ is expressed by Eq. (2) in a wide temperature range up to room temperature \cite{Kkurashima}, $n$ was estimated in the fitting range of 150--300 K for all $y$ in this study.
The hole-concentration dependence of $n$ for  Bi$_{1.74}$Pb$_{0.38}$Sr$_{1.88}$Cu$_{1-y}$Fe$_{y}$O$_{6+\delta}$ with $y = 0${--}$0.096$ is shown in Fig. 3(b), together with the preceding data of $y = 0$.~\cite{Kkurashima}
For $y = 0$, $n$ $\sim$ 4/3 corresponding to 2D ferromagnetic fluctuations.
We found that $n$ is increased by Fe substitution and $n$ is $\sim$ 5/3 at $y = 0.096$.
These results suggest that the dimensionality of the ferromagnetic fluctuations changes with the Fe substitution from 2D to 3D.

\section{Discussion}
The present results of the effects of Fe substitution in overdoped and HOD Bi$_{1.74}$Pb$_{0.38}$Sr$_{1.88}$Cu$_{1-y}$Fe$_{y}$O$_{6+\delta}$ are summarized as follows.
The temperature dependence of $\chi$ in 50 Oe on ZFC and FC exhibits hysteresis at temperatures lower than $T_{\rm sg}$, suggesting the formation of a spin-glass state related to Fe$^{3+}$ spins, and $T_{\rm sg}$ increases with the Fe substitution.
The $\chi$ in 3 T increases with the Fe substitution.
Through the fit of $\chi$ in the range of 150--300 K using the Curie--Weiss law, the Curie constant increases with the Fe substitution, and $p_{\rm eff}$ in HOD $y = 0.096$ is larger than that of a bare Fe$^{3+}$ spin.
The temperature dependence of $\rho_{\rm ab}$ revealed that the dimensionality of ferromagnetic fluctuations increases from 2D to 3D with the Fe substitution.

\begin{figure}[tbp]
\includegraphics[width=1.0\linewidth]{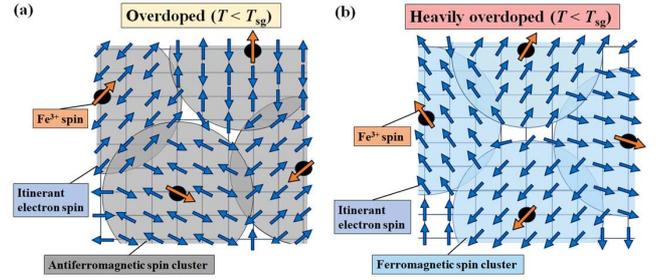}
\caption{(Color online) Schematic views of possible spin clusters below $T_{\rm sg}$ in the CuO$_{\rm 2}$ plane for (a) overdoped and (b) HOD Bi$_{1.74}$Pb$_{0.38}$Sr$_{1.88}$Cu$_{1-y}$Fe$_{y}$O$_{6+\delta}$ with $y = 0.05$. Shaded areas in (a) and (b) represent AF and ferromagnetic spin clusters, respectively. The large and small arrows represent Fe$^{3+}$ spins and itinerant electron spins, respectively.}
\label{fig:f4}
\end{figure}

In both the overdoped and HOD regimes, the values of $p_{\rm eff}$ estimated in the range of 50--300 K are smaller than that of a bare Fe$^{3+}$ spin.
The small $p_{\rm eff}$ probably originates from the formation of a spin-glass state at temperatures lower than $T_{\rm sg}$.
In contrast, the values of $p_{\rm eff}$ estimated in the range of 150--300 K exceed that of a bare Fe$^{3+}$ spin, suggesting the presence of ferromagnetic fluctuations.

As shown in Fig. 3, the exponent $n$ of $\rho_{\rm ab}$ increases with increasing $y$ from 4/3 to nearly 5/3.
According to the SCR theory \cite{Yhatatani}, $n$ = 4/3 (5/3) corresponds to 2D (3D) ferromagnetic fluctuations.
The two-dimensionality of ferromagnetic fluctuations in Fe-free Bi-2201 probably originates from the highly 2D crystal structure.
Note that 3D ferromagnetic fluctuations are proposed in La$_{2-x}$Sr$_{x}$CuO$_{4}$ with a more 3D crystal structure than that in Bi-2201.~\cite{JEsonier}
It is speculated that Fe$^{3+}$ spins stabilizing the ferromagnetic fluctuations lead to the enhancement of the spin correlation between the CuO$_2$ planes and to the 3D ferromagnetic fluctuations.

As shown in Fig. 1(b), $T_{\rm sg}$ is proportional to the Fe content, suggesting that spin clusters are formed around Fe.
In the overdoped regime, Hiraka $et \ al$. suggested the formation of AF spin clusters, on the basis of the observation of incommensurate AF peaks in the neutron scattering experiments.~\cite{Hhiraka}
Therefore, the magnetic state in the present overdoped samples would be the same as that reported by Hiraka $et \ al$.
Figure 4(a) shows a schematic view of the CuO$_2$ plane at $T$ $<$ $T_{\rm sg}$ in the overdoped regime.
The AF spin clusters represented by the shaded area are formed around Fe. 
Random orientation of Fe$^{3+}$ and itinerant electron spins between clusters due to the RKKY interaction would result in the formation of the spin-glass state.
At high temperatures, ferromagnetic fluctuations develop with the Fe substitution.
It is speculated that AF and ferromagnetic fluctuations compete with each other and that AF fluctuations surpass ferromagnetic ones at low temperatures in the overdoped regime.

In the HOD regime, the spin-glass state also appears at temperatures lower than $T_{\rm sg}$.
Our previous findings in Fe-free Bi-2201 suggest that the ferromagnetic fluctuations are more enhanced in the HOD regime than in the overdoped regime.~\cite{Kkurashima}
Moreover, recent RIXS studies of Bi-2201 have suggested that the ferromagnetic correlation developed in the HOD regime.~\cite{YYpengAF}
Therefore, ferromagnetic fluctuations would be favorable at temperatures lower than $T_{\rm sg}$ in the HOD regime.
In this case, ferromagnetic spin clusters might be formed around Fe below $T_{\rm sg}$ as shown in Fig. 4(b).
In each cluster, ferromagnetic fluctuations are stabilized owing to the Fe substitution, and the magnetic moment of each cluster is randomly oriented to form a spin-glass state.
The mechanism of the formation of a cluster spin-glass state under the ferromagnetic fluctuations is unclear.
To clarify the detailed magnetic state in Fe-substituted Bi-2201 at low temperatures, neutron-scattering experiments are being planned.

Finally, we briefly discuss the relationship of ferromagnetic fluctuations with the pseudogap.
Recently, there have been reports of various interesting phenomena occurring around the end point of the pseudogap phase in the overdoped regime of Bi-based cuprates, such as the disappearance of electronic nematicity in Bi-2212~\cite{Kishida}, the appearance of a charge order in (Bi,Pb)-2201~\cite{YYpengCONat}, and common characteristic behaviors of transport coefficients in La-substituted Bi-2201 and other cuprates~\cite{Mlizaire}.
The ferromagnetic fluctuations in La$_{2-x}$Sr$_x$CuO$_4$ were observed in the HOD regime where the pseudogap is closed~\cite{JEsonier}.
In the present Bi$_{1.74}$Pb$_{0.38}$Sr$_{1.88}$Cu$_{1-y}$Fe$_{y}$O$_{6+\delta}$ cuprates, the c-axis electrical resistivity at impurity-free $y$ = 0 in the HOD regime showed no sign of the presence of the pseudogap~\cite{Kkudo,Ykomiyama}.
Therefore, the ferromagnetic fluctuations are likely to appear outside the pseudogap phase in the phase diagram.
Future investigation of the pseudogap in Fe-substituted Bi$_{1.74}$Pb$_{0.38}$Sr$_{1.88}$Cu$_{1-y}$Fe$_{y}$O$_{6+\delta}$, where ferromagnetic fluctuations are enhanced, is expected to clarify the relationship between the pseudogap and ferromagnetic fluctuations.

\section{Summary}
In the HOD regime of Bi-2201, the Fe substitution induces the formation of a spin-glass state at temperatures lower than $T_{\rm sg}$.
The observation that $T_{\rm sg}$ increases linearly with the Fe content, that $p_{\rm eff}$ of the Fe-substituted sample is larger than that of a bare Fe$^{3+}$ spin, and that the Fe substitution changes the dimensionality of ferromagnetic fluctuations from 2D to 3D suggest that the ferromagnetic fluctuations are enhanced by magnetic Fe.
While the formation of AF spin clusters is proposed in the overdoped regime \cite{Hhiraka}, ferromagnetic spin clusters would be formed at low temperatures in the HOD regime, which should be clarified by, for example, neutron scattering experiments in the future.

\begin{acknowledgment}

We thank S. Takahashi of the Technical Division, School of Engineering, Tohoku University, for his aid in the ICP analysis.
This work was partially supported by JSPS KAKENHI (Grant No. 19H01841) from JSPS.

\end{acknowledgment}

\end{document}